\documentstyle[preprint,revtex]{aps}
\begin{document}
\draft
\preprint{Talk presented at King's College, May 1994}
\begin{title}
One Dimensional Exactly Solvable Models of\\
Strongly Correlated Electrons of $1/r^2$ Hopping and Exchange
\end{title}
\author{C. Gruber and D. F. Wang}
\begin{instit}
Institut de Physique Th\'eorique\\
\'Ecole Polytechnique F\'ed\'erale de Lausanne\\
PHB-Ecublens, CH-1015 Lausanne-Switzerland.
\end{instit}
\begin{abstract}
We review some recent progresses in study of the 1D
strongly correlated electron systems of long range hopping
and exchange. The systems are completely integrable,
with infinite number of constants of motions.
The results of the physical properties, such as
the wavefunctions, the full excitation spectrum and the
thermodynamics, are also reviewed.
\end{abstract}

\narrowtext
\section{ Introduction}
One of the main subjects of modern physics is to understand
the systems consisting of many particles interacting with
each other. Due to many degrees of freedom,
as well as due to the interactions between particles,
in general, to exactly solve their equations of motion is impossible in
most cases. However, in some very special cases,
that is, for the so-called integrable systems,
we can find exact solutions for their eigenenergies, wavefunctions
and other physical properties,
because of infinite number of symmetries of the systems.
The classical examples are dated back to the ansatz,
proposed by Bethe in 1931\cite{bethe} for the
ground state wavefunction of the spin 1/2 Heisenberg chain, as well
as the Onsager's solution for the 2D Ising model in zero
magnetic field\cite{onsager}.

In 1963, Lieb, and later
Lieb and Liniger studied the one dimensional delta potential
interacting Bose gas wavefunctions by constructing ansatz\cite{liniger},
without knowing Bethe's idea for the spin chain.
After the work of McGuire on $M=1$\cite{guire} case
and that of Flicker and Lieb\cite{lieb1} on the $M=2$ case,
Yang, Gaudin independently found the
solutions for the one dimensional delta function potential spin 1/2
fermion system for $M>2$\cite{yang,gaudin}.
Together with Baxter's solution for the statistical
mechanical system (the eight-vertex model)\cite{baxter}, factorization of
scattering matrix, often called Yang-Baxter relation, has been the key
to many exactly solvable models, both in condensed matter theory
and in high energy physics. We'd like to mention, in particular,
its great applications to the condensed matter
systems, such as the well-known Lieb-Wu solution for the
one dimensional Hubbard model\cite{wu}, the Bethe-ansatz for the
one impurity Kondo model\cite{andrei} and the Anderson model\cite{wiegmann}.

Having said those, we'd like to mention, as one exciting
progress in quantum field theory, the development of the
conformal field in two dimensions, where the infinite number
of conformal symmetries fully determine the dynamics of
the systems, since it was realized that the scaling invariant
systems also respect the conformal symmetries\cite{confor}. The conformal
field theory has become a powerful tool for the critical
phenomenon theory, where the fluctuations on all length scales
are important and the systems are scale invariant.
Generally speaking, such infinite number of symmetries determining
systems' dynamics may also be considered as one example of
solvability when many degrees of freedom are involved.
In general, exact solvability of many particle systems
( in field theory case, we have continuum of particles) have
been attracting attentions from both condensed matter physicists,
and mathematical physicists, as well as from field theorists.

In this talk, we'd like to view some recent work
in the approach to the 1D strongly correlated electron systems
by solving them exactly. There has been
considerable general interest in study of low dimensional
strongly correlated electrons, largely due to the discovery
of high temperature superconductivity\cite{anderson}.
The two dimensional t-J model and the Hubbard model,
believed to model the high
temperature superconductors, have been studied intensively
through various methods\cite{su}. The fixed point
behaviors of the systems from one dimension to two dimensions
have been one of the major interests of the theorists working in
the field. Whether the normal state of
the new 2D-like superconducting materials is a non-fermi liquid,
is one of main focuses for its theoretical study.

The ordinary one dimensional t-J model with supersymmetry\cite{sakar},
and the one dimensional Hubbard model
are exactly solvable\cite{wu}, with the technique of the Bethe-ansatz.
Different from the Bethe-integrable systems like these, consisting of
one type of integrable family, Calogero and Sutherland
have developed another quite interesting integrable
interacting many particle family with
long range $1/r^2$ interactions\cite{calo,suth72}.
The interests in the Calogero-Sutherland type systems are
renewed after Haldane and Shastry's independent works
on the integrable one dimensional spin systems with
long range exchange interaction in 1988\cite{shas88,hald91}.
The close relation of the Calogero-Sutherland
type system with quantum chaos, random matrices, has been greatly
acknowldeged. In our study, however, we focus on the integrable
correlated electron systems in one dimension related to the
Calogero-Sutherland-type quantum system, that is, the one dimensional
t-J models of $1/r^2$ hopping and exchange.

\section{1D t-J model on uniform lattice}

Let us first consider the $1/r^2$ supersymmetric t-J model
on a one dimensional lattice of equal spaced sites.
The Hamiltonian for the system is given by
\begin{equation}
H= (1/2) P_{G}
\left[ -\sum_{1\le i\ne j \le L}\sum_{\sigma=1}^{N} t_{ij} ( c_{i\sigma}
^{\dagger}
c_{j\sigma}) +
\sum_{1\le i\ne j \le L}
J_{ij} \left[ P_{ij}-(1-n_i)(1-n_j)\right] \right] P_{G},
\label{eq:original}
\end{equation}
where we take the hopping matrix and the spin exchange interaction to be
$t_{ij}/2 = J_{ij} = 1/d^2(i-j)$, and $d(n)=(L/\pi) \sin (n\pi/L)$
is the chord distance, with $L$ the size of the lattice.
The operator $c_{i\sigma}^{\dagger}$ is the
fermionic operator to create an electron with spin component $\sigma$
at site $i$, $c_{i\sigma}$ is the corresponding
fermionic annihilation operator. Their anti-commutation relations
are given by $\{c_{i\sigma_i},c_{j\sigma_j}^{\dagger}\}_+
=\delta_{ij} \delta_{\sigma_i\sigma_j},
\{c_{i\sigma_i},c_{j\sigma_j}\}_+=0, \{c_{i\sigma_i}^\dagger
,c_{j\sigma_j}^\dagger\}_+=0$.
We assume that the spin component
$\sigma$ takes values from $1$ to $N$.
The Gutzwiller projection operator $P_{G}$
projects out all the double or multiple occupancies,
$P_{G} = \prod_{i=1}^L P_{G}(i)$, and $P_{G}(i) =\delta_{0,n_i}
+\delta_{1,n_i}$, with $n_i=\sum_{\sigma=1}^N c_{i\sigma}^\dagger
c_{i\sigma}$.
The operator $P_{ij}=\sum_{\sigma=1}^N \sum_{\sigma'=1}^N
c_{i\sigma}^{\dagger} c_{i\sigma'} c_{j\sigma'}^{\dagger} c_{j\sigma}$
exchanges the spins of the electrons at sites $i$ and $j$, if both sites
are occupied. $n_i$ and $n_j$ are the electron number operators
at sites $i$ and $j$.

The $SU(2)$ model was first introduced by Kuramoto and Yokoyama,
who provided the ground state of the system, the Gutzwiller
projected Fermi sea away from half-filling\cite{kura91}.
More generalized Jastrow wavefunctions, including the ground state
as a member, were also introduced\cite{wang,ha}.
The asymptotic energy spectrum was derived by Kawakami based
on the assumption of asymptotic scattering
matrix factorization\cite{kawakami92}.
The system is identified as
a free Luttinger liquid\cite{kura91,kawakami92,wang,ha}.
In particular, for the $SU(2)$ case, the asymptotic Bethe-ansatz
spectrum was explicitly shown to be exact, and the correct
thermodynamics was given when the spinon rotation was properly
taken into account\cite{wang}, in terms of a set of generalized
Laughlin-Jastrow wavefunctions. In general,
exact solvability implies existence
of infinite number of constants of motion.
At half-filling, this system reduces to the Haldane-Shastry spin.
Quite recently, the constants of motion of the spin chain have been
studied\cite{fowler}, using the operators $M_{ij}$\cite{poly}.
In our most recent work, we have pointed out that their results
can apply to the case where the holes are involved, with the superalgebra
representation of the fermionic fields
in the t-J model\cite{gruber1}, by mapping
the system to a mixture gas of fermions and bosons on the one dimensional
lattice.

\subsection{ The constants of motion}

For the t-J model, we may introduce two new fields, the $f$ and
$b$ fields.
For the new fields, we have $\{f_{i\sigma},f_{j\sigma'}\}_{+} =0,
\{f_{i\sigma}^{\dagger}, f_{j\sigma'} \}_{+} =
\delta_{ij} \delta_{\sigma\sigma'}, \left[ b_i, b_j \right]=0,
\left[ b_i, b_j^\dagger \right] =\delta_{ij} $. The $b$ field
always commute with the $f$ field. The size of the Hilbert space at
each site is $\infty$ in this case.
After projecting out the zero occupancy, double and multiple
occupancies, the new Hilbert space is equivalent to the one
defined by the $c$ fields with no double or multiple occupancies.
In particular, we may represent the fermionic electron operators
$c_{i\sigma}^{\dagger}$ and $c_{i\sigma}$ as:
\begin{eqnarray}
&& P_{G}(i) c_{ i \sigma}^{\dagger} P_{G}(i) =
\delta_{1,n_b^i+n_f^i} f_{i \sigma}^{\dagger} b_i \delta_{1,n_b^i+n_f^i}
\nonumber\\
&& P_{G}(i) c_{i\sigma} P_{G}(i)
= \delta_{1,n_b^i+n_f^i} b_i^{\dagger} f_{i\sigma}
\delta_{1,n_b^i+n_f^i},
\label{eq:mapping}
\end{eqnarray}
where $n_b^i +n_f^i = b_i^{\dagger} b_i +
\sum_{\sigma =1}^{N} f_{i\sigma}^{\dagger} f_{i\sigma}$.
Hence, a state vector can be written as
\begin{eqnarray}
|\phi> = \sum_{\sigma_1, \sigma_2, \cdots, \sigma_{N_e} }
&&\sum_{ \{x\},\{y\} } \phi ( x_1\sigma_1, x_2\sigma_2,
\cdots, x_{N_e}\sigma_{N_e} | y_1, y_2, \cdots, y_Q )\times \nonumber\\
\times &&f_{x_1\sigma_1}^{\dagger} f_{x_2\sigma_2}^{\dagger} \cdots
f_{x_{N_e} \sigma_{N_e}}^{\dagger} b_{y_1}^{\dagger} b_{y_2}^{\dagger}
\cdots b_{y_Q}^{\dagger} |0>,
\label{eq:amp}
\end{eqnarray}
where $N_e$ is the number of $f$ fermions on the lattice, $Q$ is the
number of $b$ bosons, and we require that
$x_i\ne x_j \ne y_k \ne y_l$, and that
the $f$ fermion positions $\{x\}$ and the $b$
boson positions $\{y\}$ span the whole chain.
Obviously, $N_e$ is also the number of
the electrons, and $Q$ is also the number of holes on the lattice.
The amplitude $\phi$ is anti-symmetric when exchanging
$(x_i\sigma_i)$ and $ (x_j\sigma_j)$, and symmetric in the
boson coordinates $\{y\} = (y_1, y_2, \cdots, y_Q)$.
The Hamiltonian of the supersymmetric t-J model, using the
mapping Eq.~(\ref{eq:mapping}) in a straightforward way, can be
written in terms of
the fermionic $f$ field and the bosonic $b$ field.

With the above mapping, we may write the eigen-energy equation of the
supersymmetric t-J model in first quantized form.
Following Ref.~(\cite{poly}), we define the ``exchange operator'' $M_{ij}$ as
$M_{ij} F(q_1,q_2,\cdots, q_i, \cdots, q_j, \cdots, q_L )
= F(q_1,q_2, \cdots, q_j, \cdots, q_i, \cdots, q_L)$,
where the function $F$ is an arbitrary function of some position variables
$(q_1, q_2, \cdots, q_L)$, i.e., the operator $M_{ij}$
exchanges the positions $q_i, q_j$
of the particles $i $ and $j $.
In terms of such exchange operators, the eigen-energy equation of the
t-J model takes the form as follows
\begin{equation}
-(1/2) \left[ \sum_{1\le i\ne j\le L} d^{-2} (q_i -q_j)
M_{ij} \right]
\phi(\{q\};\{\sigma\} ) = E \phi(\{q\};\{\sigma\}),
\label{eq:eigen}
\end{equation}
where $ \{ q \}=(q_1, q_2, \cdots, q_L) =
(x_1, x_2, \cdots, x_{N_e}, y_1, y_2, \cdots, y_Q)$,
and $\phi(\{q\};\{\sigma\})
= \phi(q_1\sigma_1,q_2\sigma_2,\cdots,q_{N_e}\sigma_{N_e}
|q_{N_e+1} q_{N_e+2} \cdots q_{L})=\phi (x_1\sigma_1, x_2\sigma_2,
\cdots, x_{N_e}\sigma_{N_e} | y_1, y_2, \cdots, y_Q)$ is the
amplitude of the state vector of Eq.~(\ref{eq:amp}).
$\{\sigma\} = (\sigma_1, \sigma_2, \cdots, \sigma_{N_e})$
are the spin variables of the $f$ fermions.
The operation $M_{ij}$ is defined in the conventional way:
$M_{ij} \phi(\{q\};\{\sigma\})
=\phi (\{q'\};\{\sigma\}) $, with $\{q\}=(q_1,q_2,\cdots,q_i,\cdots,q_j,
\cdots, q_L)$ and $\{q'\}=(q_1,q_2,\cdots,q_j,\cdots,q_i,\cdots,q_L)$.
Therefore, the original t-J model is written as an eigenvalue
problem for a mixture of the $f$ fermions and the spinless
$b$ bosons in terms of the position exchange operator $M_{ij}$.

The constants of motion have been studied recently for the spin chain
of $1/r^2$ exchange interaction. In the work, a
system of identical bosons on the chain is considered,
and the wavefunctions is totally symmetric when interchanging
two particles.
In terms of the operator $M_{ij}$, the following commutation
results have been found\cite{fowler}
\begin{equation}
\left[ I_n, I_m \right] = 0,
\end{equation}
where $I_n = \sum_{i=1}^L \pi_i^n $,
with $\pi_i = \sum_{ j(\ne i)} (z_j/z_{ij}) M_{ij} $,
$z_i = e^{ 2\pi i q_i /L}, z_{ij} = z_i -z_j$,
and $n, m =0, 1, 2, \cdots, \infty$.
It was found that
all these quantities commute with the
Hamiltonian $H = \sum_{1\le i\ne j \le L} |z_i -z_j|^{-2} M_{ij}$
as long as the particles occupy the whole chain.
For a system of identical bosons on the chain,
the wavefunction is totally symmetric when we simultaneously interchange
spins and positions of two particles. The effect of the exchange
operator $M_{ij}$ is just equivalent to the effect of
the spin exchange operator alone. With this method,
the constants of motion for the $SU(N)$ Haldane-Shastry spin chains
were constructed in Ref.\cite{fowler}.

In the recent work\cite{gruber1},
we have stressed that, in the language of the
exchange operators $M_{ij}$, the commutation
results proved by Fowler and Minahan
hold as operator identities. The central issue is that the forms
of wavefunctions of many particle systems, as well as the statistics of the
particles or the types of the particles, do $\it not $ matter in order
for the commutators to hold, as long as the particles occupy the whole
chain. We may then apply the ``exchange operator formalism''
to the wavefunctions of mixtures of fermions and bosons.
Therefore, from the eigen-equation Eq.~(\ref{eq:eigen}),
we thus conclude that in the first quantization,
all the invariants of the t-J model are the same
$I_n$'s as in Ref.\cite{fowler}.

With the permutation properties of the amplitude $\phi$ for the mixture of
bosons and fermions, it is straightforward to
write all the invariants of motion of the t-J model in the
second quantization form using the $I_n$'s.
For instance, the exchange operation between the $f$ fermion positions
is equivalent to the spin exchange operation ( minus sign involved ),
the exchange operation between $b$ boson positions is equivalent to
the hole-hole interaction term, and exchange operation between
$f$ fermion and $b$ boson positions is equivalent to the
electron hopping. Such procedure to reduce an $I_n$ to a second
quantized form is quite simple, and
we do not write all the details.
Thus we provide a systematical way to construct all
the invariants of motion for the $1/r^2$
supersymmetric t-J model,
either in first quantized or in second quantized forms.

\subsection{The Jastrow product wavefunctions}

\def\hh{{\sl \Phi_{\vphantom{h}s}}}
\def\ss{{\sl \Phi_h}}
Having said the integrability of the t-J model, we'd like to mention
that the exact eigenwavefunctions can be constructed explicitly
for the system.
In the $SU(2)$ case,
states in the Hilbert space can be represented by spin and hole
excitations from the fully-polarized
up-spin state $|P>$.
If we let $Q$ denote the number of holes and $M$ denote the number of
down-spins, then $S_z$ is given by $S_z=(N-Q)/2-M$.  The wavefunctions
are given by
\begin{equation}
|\psi\rangle=\sum_{x,y}\psi(x,y)
\prod_\alpha S^{-}_{x_\alpha} \prod_i h_{y_i}^\dagger
|P\rangle,
\label{eq:wave}
\end{equation}
where the amplitude $\psi(x,y)$ is symmetric in
$x \equiv (x_1,x_2,\ldots,x_M)$, the positions of the down-spins,
and antisymmetric in $y \equiv ( y_1,y_2,\ldots,y_Q)$,
the positions of the holes.
$S^{-}_{x_{\alpha}}=c_{x_\alpha\downarrow}^\dagger
c_{x_\alpha\uparrow}^{\vphantom{\dagger}}$ is the spin-lowering operator
at site $x_\alpha$ and $h_{y_i}^\dagger=c_{y_{i\uparrow}}$ creates a hole at
site $y_i$.

We can construct a general class of states corresponding to states of
uniform motion and spin polarization, by
generalizing Kuramoto and Yokoyama's Jastrow ground-state
as follows\cite{wang}
\widetext
\begin{eqnarray}
\psi_G(x,y;J_s,J_h)=&&\exp\left[{2\pi i\over N}
\left({J_s\sum_\alpha x_\alpha}+{J_h\sum_i y_i}\right)\right]{\bf
 \Psi}_0(x,y)\nonumber\\
\quad {\bf \Psi}_0(x,y)=&&
\prod_{\alpha<\beta}d^2(x_\alpha-x_\beta)\prod_{i<j}d(y_i-y_j)
\prod_{\alpha,i}d(x_\alpha-y_i).
\label{eq:gutz}
\end{eqnarray}
\narrowtext
Here,  $J_s$ and $J_h$ govern the (uniform) momenta of
down-spins and holes respectively.  $J_s$ and $J_h$ take on either
integral or half-integral values as appropriate to insure that
$\psi_G$ has the correct periodicities under $x_\alpha \to x_\alpha+N$
and $y_i \to y_i+N$.

The wavefunction $\psi_G$ is found to
be an exact eigenstate of $H$ with energy\cite{wang}
\begin{eqnarray}
{N^2\over\pi^2t}E=&&{2\over3}M(M^2-1)-2MJ_s(N-J_s)\nonumber\\
&&+Q\biggl[{1\over3}(N^2-1)+{2\over3}(Q^2-1)+{1\over2}(M+Q)(2M-Q)\nonumber\\
&&\quad\quad-2J_h(N-J_h)+2(J_s-J_h)^2\biggr]
+{1\over48}(N^2-1)(N-Q).
\end{eqnarray}
which is valid under the conditions
$|J_s-N/2|\le N/2-(M-1+Q/2)$,
$|J_h-N/2|\le N/2-(M+Q-1)/2$ and
$|J_h-J_s|\le(M+1)/2$.
For a given $S_z$, the minimum energy is given when $J_s$ and $J_h$
are as close to $N/2$ as possible.
When $Q=0$ this reduces to the result for the
Heisenberg chain\cite{shas88,hald91}.
These energy levels were also found independently in Ref.\cite{ha}.
The equivalence of the two representations
is clear when identifying
$J_{\uparrow} = J_h -N+(M+Q+1)/2$ and $J_{\downarrow} =
J_h - J_s - (M-1)/2$. Besides these explicitly constructed excited
wavefunctions, we'd like to emphasize that the full spectrum
and the thermodynamics can also be written out in terms of
Jastrow wavefunctions of more generalized form\cite{wang}.

\section{The 1D t-J model on nonuniform lattice}

\subsection{The integrability}

Besides the above integrable t-J model on equally-spaced sites,
let us consider another supersymmetric
t-J model of $1/r^2$ hopping and exchange
on a chain with sites not equally spaced.
The positions of the sites $x_1, x_2, \cdots, x_{L}$ are determined
by the equation
\begin{equation}
x_i = \sum_{1\le j (\ne i)\le L} 2/(x_i-x_j)^3.
\label{eq:positions}
\end{equation}
An integrable spin chain was introduced on a non-uniform lattice
on which the sites are positioned by the equation\cite{poly}.
Later, Frahm realized that the solutions of the equations are
the zeros of the Hermit polynomials $H_L(x)$\cite{frahm}.
The Hermite polynomial is defined as
\begin{equation}
H_L(x)=(-1)^L  e^{x^2} {\partial^L e^{-x^2} \over \partial x^L}.
\end{equation}
It was well-known that the Hermit polynomial has $L$ roots, all of
which are real numbers, and the lattice is thus well defined.

Doping the spin chain, we have introduced the following supersymmetric
t-J model\cite{gruber1}
\begin{equation}
H=(1/2) P_{G} \left[
-\sum_{1\le i\ne j \le L} \sum_{\sigma=1}^{N} t_{ij} ( c_{i\sigma}
^{\dagger}
c_{j\sigma}) +\sum_{1\le i\ne j \le L}
J_{ij} \left[ P_{ij}-(1-n_i)(1-n_j)\right] \right] P_{G},
\end{equation}
where the hopping matrix and the anti-ferromagnetic
exchange interaction are given by $t_{ij}/2 = J_{ij} = 1/(x_i-x_j)^2$, and
each site is occupied at most by one electron.

In the half filling case $N_e = L$, this system reduces to the spin chain
that has been studied before,
for which a
similar exchange operator formalism has been developed.
Let us just write down the results obtained in Ref.~{\cite{poly}}:
$[I_n, I_m] = 0, [I_n, H]=0$, where $I_n = \sum_{i=1}^L h_i^n$,
$h_i=a_i^{\dagger} a_i$, and
$a_i^{\dagger} = \pi_i^{\dagger} + i q_i, a_i=\pi_i -iq_i$, with
$\pi_i =\sum_{j(\ne i)} i(q_i-q_j)^{-1} M_{ij}, H=\sum_{i\ne j}
(q_i-q_j)^{-2} M_{ij} $, and $n, m=0,1,2,\cdots, \infty$.
Here, all the particles are put on the chain where the sites
are positioned as determined by the Eq.~(\ref{eq:positions}).
We may relate the operation of exchanging particle positions
to the operation to exchange particle spins, by assuming
that we have identical bosons again, for which the wavefunctions
are totally symmetric when we exchange the spins and positions of
two particles simultaneously. With this assumption,
from $I_n$'s, we thus can derive all the invariants of motion
written in terms of the spin exchange operators alone.

Again, all commutation results written in terms of the exchange operators
$M_{ij}$ obtained by Polychronakos
hold as operator identities, as long as the particles occupy the whole chain
of the sites positioned in the special way.
The forms of the wavefunctions
do ${\it not}$ matter at all. Thus the commutation results
may apply to wavefunctions of
particles of arbitrary statistics or wavefunctions of
mixtures of particles of different statistics on the chain.
Mapping our new supersymmetric t-J model
in terms of the $b$ and $f$ fields,
we may also write the eigen-energy equation in
first quantized form. In terms of the exchange operators between
the positions of the bosons and fermions, the Hamiltonian takes
the form
\begin{equation}
H=(-1/2) \sum_{1\le i\ne j \le L} (q_i-q_j)^{-2} M_{ij}.
\label{eq:eigen2}
\end{equation}
Applying the formalism to this t-J model, in a similar way
we may obtain all the invariants, either in first quantized or
in the second quantized form, which commute among themselves and
with the Hamiltonian. Thus this supersymmetric t-J model is
also completely integrable.

\subsection{The wavefunctions}

The Hamiltonian commutes with the permutation operator
$T_{ij} = P_{ij}^{\sigma} M_{ij}$ exchanging the $f$ fermion spin and
position simultaneously.
Let us work in the Hilbert space where the number of
fermions of each flavor is fixed,
i.e. $\tilde N_{\sigma}$, $\sigma =1, 2, \dots, N$, is fixed.
Gruber and Wang have considered
the following wavefunction in Jastrow product form\cite{gruber2},
\begin{equation}
\phi (x_1\sigma_1, x_2\sigma_2, \cdots, x_{N_e}\sigma_{N_e}
|y_1, y_2, \cdots, y_Q)
=\prod_{i<j} (x_i-x_j)^{\delta_{\sigma_i\sigma_j}} e^{i{\pi\over 2}
sgn (\sigma_i -\sigma_j)},
\label{eq:groundstate}
\end{equation}
where $\{x\}$ and $\{y\}$ span the whole lattice.

It was shown that the wavefunction $\phi$ is an eigenstate
with eigenvalue\cite{gruber2}
\begin{equation}
E = -L(L-1)/4 + (1/2) \sum_{\sigma =1}^N (\tilde N_\sigma -1) \tilde N_\sigma.
\label{eq:eigenvalue1}
\end{equation}
Although it is expected that this wavefunction is the
lowest energy state in the subspace of fixed
$\tilde N_1, \tilde N_2, \cdots, \tilde N_N$, we were not able
to prove it. However, in the case of $SU(2)$, the small lattice
diagonalization up to 8 sites confirms this conclusion.

For fixed number $N_e$ of the electron number, the minimum of the
energy is obtained when $|\tilde N_\sigma - \tilde N_{\sigma'}|$
is as small as possible for each pair $\sigma \ne \sigma'$.
In the $SU(2)$ case, the above result becomes
\begin{equation}
E = (-1)L(L-1)/4 + (1/2) \tilde N_{\uparrow}  ( \tilde N_{\uparrow} -1)
+(1/2) \tilde N_{\downarrow} (\tilde N_{\downarrow}-1),
\label{eq:energy}
\end{equation}
where $\tilde N_{\uparrow}$ and $\tilde N_{\downarrow}$ are the numbers of
the up-spin electrons and the down-spin electrons respectively.
For fixed number of electrons on the chain, i.e. for fixed $N_e$,
the minimum of the energy given in Eq.~(\ref{eq:energy})
is obtained when $S_z=0$ for even $N_e$, or when
$S_z=\pm 1/2$ for odd $N_e$.
Therefore, the ground state
is a spin singlet ( respectively spin 1/2 ) state for even ( respectively odd)
number of electrons on the chain.
In particular, for an even number of electrons on the chain,
the ground state energy is
\begin{equation}
E_G= (-1/4) L(L-1) + N_e^2/4 -N_e/2,
\end{equation}
while for an odd number of electrons it is
\begin{equation}
E_G = (-1/4)L(L-1) +({N_e -1 \over 2})^2.
\end{equation}
The charge susceptibility of the ground state $\chi_c$ is given by
$\chi_c^{-1} =\partial^2  E_G/\partial N_e^2 = 1/2$, independent of the
electron concentration.
The charge susceptibility is also finite at
half-filling $N_e=L$, in spite of the existence of
a metal-insulator phase transition at half-filling for this system.

To study the spectrum of the system away from half-filling,
we follow the idea introduced in Ref.~\cite{poly}, defining the operators:
\begin{eqnarray}
&&\pi_j = i \sum_{k(\ne j)} (q_j-q_k)^{-1} M_{jk} = \pi_j^\dagger ,\nonumber\\
&&a_j^\dagger = \pi_j +i q_j,\nonumber\\
&&a_j = \pi_j -iq_j,
\end{eqnarray}
which satisfy the following commutation relations:
\begin{eqnarray}
&&[ \pi_j, \pi_k ] =0 \nonumber\\
&&[q_j, H] = i\pi_j\nonumber\\
&&[\pi_j, H] = -2 i \sum_{k(\ne j)} (q_j-q_k)^{-3}.
\end{eqnarray}
Then using the property of the roots of the Hermite polynomial
we have
\begin{equation}
[\pi_j, H]=-iq_j,
\end{equation}
and
\begin{eqnarray}
&& [ a_j^\dagger, H] = -a_j^\dagger\nonumber\\
&& [ a_j, H] =a_j.
\end{eqnarray}
Therefore the operators
$A_i^\dagger (\nu) = a_i^\dagger S_i^{(\nu)}, i=1, 2, \cdots, N_e$,
where $\nu =0,\pm, z$ for the $SU(2)$ case with $S_i^{(0)}=1$,
will act as raising operators,
while their hermitian conjugate $A_i(\nu)$
will act as lowering operators. It thus follows that the wavefunction
\begin{equation}
\phi_{\{n\}, \{\nu\}} = \sum_{P}
\prod_{i=1}^{N_e} (A_i^\dagger(\nu_i))^{n_{P_i}}\phi
\end{equation}
with $\{n\}=(n_1, n_2, \cdots, n_{N_e}),
n_i\ge 0, \{\nu\}=(\nu_1, \nu_2, \cdots,
\nu_{N_e} )$, is either an eigenstate with energy
\begin{equation}
E_{\{n\}} = E +\sum_{i=1}^{N_e} n_i
\end{equation}
or zero.
Moreover, for the $b$ boson degrees of freedom, we
have following property\cite{gruber2}
\begin{equation}
a_i \phi =0, i\in (N_e +1, N_e +2, \cdots, L),
\label{eq:holes}
\end{equation}
We have also shown the following results are true:
\begin{eqnarray}
\sum_{\alpha =N_e +1}^L &&a_{\alpha} \phi =0,\nonumber\\
\sum_{i=1}^{N_e} &&a_i \phi =0.
\end{eqnarray}
Furthermore, we realize that
\begin{equation}
[\sum_{i=1}^{N_e} A_i(z) ] \phi =0.
\label{eq:anni}
\end{equation}
In the particular case where $\tilde N_{\uparrow} = \tilde N_{\downarrow}$,
the wavefunction function $\phi$ is a spin singlet and
we may globally rotate Eq.~(\ref{eq:anni}) in the
spin space, giving us\cite{gruber2}
\begin{equation}
\left[ \sum_{i=1}^{N_e} A_i(\pm) \right] \phi = 0.
\end{equation}
These results have demonstrated the impossibilities of
constructing non-vanishing eigenstates with these lowering operators
and the wavefunctions $\phi$ in the subspace
where the number of electrons of each flavor is fixed.
The results are in support of our conclusion that in general
the Jastrow wavefunction is the lowest energy state in the subspace
of fixed number of particles of each flavor.

In the end, we have concluded that the excitation
spectrum of the system is of the form\cite{gruber2}
\begin{eqnarray}
E(s) = &&E + s; \nonumber\\
&&s\in (0, 1, 2, \cdots, s_{max}),
\label{eq:fullspectrum}
\end{eqnarray}
i. e., the spectrum of this t-J model
consists of equal-spaced energy levels. Since the model is on
a finite chain, $s_{max}$ is finite. In the special case
of $SU(2)$, the small lattice diagonalization up to 8 sites
suggests that the highest energy level is given by
\begin{equation}
E_{max} (Q) = L(L-1)/4 -Q(Q-1)/2,
\end{equation}
where $Q$ is the number of holes on the chain.
The feature of the t-J model spectrum consisting of equal-distant
energy levels may also be seen by taking the strong
interaction limit of the Calogero-Sutherland-Morse quantum system
for a mixture of fermions and bosons,
\begin{equation}
H = (-1/2) \sum_{i=1}^L \partial^2/\partial q_i^2 +
\sum_{i=1}^L l^2 q_i^2 /2 +
\sum_{i<j} l(l-M_{ij})/(q_i-q_j)^2,
\end{equation}
where there are $N_e$ fermions with spins and $Q$ spinless bosons,
$M_{ij}$ permutes the positions of the particles $i$ and $j$ only.

Finally, we'd like to point out that the Hilbert space of the t-J model
has very interesting degeneracy pattern, as indicated
by the finite lattice study. It still remains to find a systematic
rule to characterize the degeneracy of each energy level.
So does the thermodynamics remain to be solved.

\section{Concluding Remarks}

In this talk, we have reviewed some recent progresses in
study of the 1D strongly correlated electron systems.
They are exactly solvable systems, close related to the
continuum Calogero-Sutherland quantum nonrelativistic system
with long range pairwise interaction. There have still been
constant interests in such systems recently.
In particular, we would like to mention that the Calogero-Sutherland
system can be interpreted,
with Haldane's new definition of statistics based on counting
state number, as a system of particles obeying
fractional statistics, where the dynamic coupling is closely
related to the fractional statistical parameter\cite{benard,li,muthry}.
We also wish to mention that the complete analytical confirmation
for the energy spectrum
and thermodynamics of the
$1/r$ Hubbard model with finite $U$, proposed
by Gebhard and Ruckenstein have been still lacking\cite{rucken92}.
In the strong coupling limit $U=\infty$, the Gutzwiller-Jastrow
wavefunctions were demonstrated to be eigenfunctions of the Hamintonian,
the spectrum conjectured by Gebhard and Ruckensten was explicitly
proved to be exact\cite{coleman,wang3}.
It remains to find the wavefunctions and the constants of motion
for the finite $U$ case for this Hubbard model with
only relativistic right movers.
Further exciting results are expected for these
interesting integrable systems from the communities
of condensed matter theory, high energy physics, as
well as of mathematical physics.



\begin{references}
\bibitem{bethe} H. Bethe, Z. Phys. {\bf 75}, 205 (1931).
\bibitem{onsager} L. Onsager, Phys. Rev. {\bf 65}, 117 (1944).
\bibitem{liniger} E. H. Lieb and Liniger, Phys. Rev. {\bf 130}, 1605 (1963).
E. H. Lieb, Phys. Rev. {\bf 130}, 1616 (1963).
\bibitem{guire} J. B. Guire, J. Math. Phys. {\bf 6}, 432 (1965).
\bibitem{lieb1} F. Flicker and E. H. Lieb, Phys. Rev. {\bf 161}, 179 (1967).
\bibitem{yang} C. N. Yang, Phys. Rev. Lett. {\bf 19}, 1312 (1967).
\bibitem{gaudin} M. Gaudin, Phys. Lett. {\bf A24}, 55 (1967).
\bibitem{baxter} R. J. Baxter, Ann. of Phys. {\bf 70}, 193 (1972).
\bibitem{wu} E. H. Lieb and F. Y. Wu, Phys. Rev. Lett. {\bf 20},
1445 (1968).
\bibitem{andrei} N. Andrei, Phys. Rev. Lett. {\bf 45}, 379 (1980),
N. Andrei, K. Furuya and J. H. Lowenstein, Rev. Mod. Phys.
{\bf 55}, 331 (1983).
\bibitem{wiegmann} P. B. Wiegmann, J. Phys. {\bf C 14}, 1463 (1981).
\bibitem{confor}A. A. Belavin, A. M. Polyakov and
A. B. Zamolodchikov, Nucl. Phys. {\bf B 241}, 333 (1984).
\bibitem{anderson} P. W. Anderson, Science {\bf 235}, 1196 (1987).
\bibitem{su} See review article by Z. B. Su and L. Yu ( preprint of
ICTP, Trieste), (1993), detailed references therein.
\bibitem{sakar} S. Sakar, J. Phys. {\bf A23}, L 409 (1990),
{\bf 24}, 1137 (1991), {\bf 24}, 5775 (1991).
P. A. Bares, G. Blatter and M. Ogata,
Phys. Rev. {\bf B 44}, 130 (1991).
\bibitem{shas88} B. S. Shastry, Phys. Rev. Lett. {\bf 60}, 639 (1988);
Phys. Rev. Lett. {\bf 69}, 164 (1992).
\bibitem{hald91} F. D. M. Haldane, Phys. Rev. Lett. {\bf 60}, 635 (1988);
Phys. Rev. Lett. {\bf 66}, 1529 (1991).
\bibitem{shastry} B. S. Shastry and B. Sutherland,
Phys. Rev. Lett. {\bf 70}, 4092 (1993).
\bibitem{shastry2} B. Sutherland and B. S. Shastry,
Phys. Rev. Lett. {\bf 75}, 5 (1993).
\bibitem{vacek} K. Vacek, A.
Okiji and N. Kawakami, Osaka-Kyoto Preprint, 1993.
\bibitem{kura91} Y. Kuramoto and H. Yokoyama, Phys. Rev. Lett. {\bf
67}, 1338 (1991).
\bibitem{kawakami92} N. Kawakami, Phys. Rev. B {\bf 45}, 7525 (1992).
\bibitem{kawakami93} N. Kawakami, Phys. Rev. {\bf B 46}, 1005 (1992).
\bibitem{gruber1} D. F. Wang and C. Gruber,
IPT-EPFL preprint (November 1, 1993),
"Invariants of the 1D $1/r^2$ t-J Models".
\bibitem{gruber2} C. Gruber and D. F. Wang, IPT-EPFL preprint (January 1994),
"Exact Results of the 1D $1/r^2$ t-J Model
without Translational Invariants".
\bibitem{wang} D. F. Wang, James T. Liu and P. Coleman,
Phys. Rev. B {\bf 46}, 6639 (1992).
\bibitem{poly} A. P. Polychronakos, Phys. Rev. Lett. {\bf 69}, 703 (1992),
Phys. Rev. Lett. {\bf 70},
2329 (1993).
\bibitem{fowler} M. Fowler and J. A. Minahan, Phys. Rev. Lett.
{\bf 70}, 2325 (1993).
\bibitem{frahm} Holger Frahm, J. Phys. A. {\bf 26}, 473 (1993).
\bibitem{calo} F. Calogero, J. Math. Phys. {\bf 10}, 2191 (1969).
\bibitem{suth72} B. Sutherland, Phys. Rev. A {\bf 5}, 1372 (1972),
Phys. Rev. A {\bf 4}, 2019 (1971), J. Math. Phys. {\bf 12}, 251 (1971),
J. Math. Phys. {\bf 12}, 246 (1971),
\bibitem{rucken92} F. Gebhard and A. E. Ruckenstein, Phys.
Rev. Lett. {\bf 68}, 244 (1992).
\bibitem{ha} Z. Ha and Haldane, Phys. Rev. {\bf B 46}, 9359 (1993).
\bibitem{coleman} D. F. Wang, Q. F. Zhong and P . Coleman,
Phys. Rev. {\bf B 48}, 8476 (1993).
\bibitem{wang3} D. F. Wang, Phys. Rev. {\bf B 48}, 10556 (1993).
\bibitem{benard} D. Benard and Y. S. Wu, cond-mat preprint April (1994).
\bibitem{li} S. Li etc, Macmaster Univ., cond-mat preprint April (1994).
\bibitem{muthry} Muthry, cond-mat preprint April (1994).
\end{references}
\end{document}